\documentclass[conference,letterpaper]{IEEEtran}
\IEEEoverridecommandlockouts
\usepackage{cite}
\usepackage{amsmath,amssymb,amsfonts}
\usepackage{algorithmic}
\usepackage{graphicx}
\usepackage{textcomp}
\usepackage{xcolor}
\def\BibTeX{{\rm B\kern-.05em{\sc i\kern-.025em b}\kern-.08em
    T\kern-.1667em\lower.7ex\hbox{E}\kern-.125emX}}
\usepackage{verbatim}
\usepackage{graphicx}
\usepackage{array}
\usepackage{makecell}
\usepackage{threeparttable}
\usepackage{caption}
\usepackage{subfigure}
\usepackage{enumerate}
\usepackage{setspace}
\usepackage{url}
\usepackage{color}
\usepackage{multirow}
\usepackage{latexsym}
\usepackage{listings}
\usepackage{xcolor}
\begin{document}

\title{An efficient and secure scheme of verifiable computation for Intel SGX\\
}


\author{\IEEEauthorblockN{Wenxiu Ding\IEEEauthorrefmark{1}, Wei Sun\IEEEauthorrefmark{1}, Zheng Yan\IEEEauthorrefmark{1}\IEEEauthorrefmark{2}, and Robert H. Deng\IEEEauthorrefmark{3}}
\IEEEauthorblockA{\IEEEauthorrefmark{1}School of Cyber Engineering, Xidian University, Xi'an, Shaanxi, China}
\IEEEauthorblockA{\IEEEauthorrefmark{2}Department of Communications and Networking, Aalto University, Espoo, Finland}
\IEEEauthorblockA{\IEEEauthorrefmark{3}School of Information System, Singapore Management University, Singapore	}
Emails: {wxding@xidian.edu.cn, zhgsunwei12188@126.com, zheng.yan@aalto.fi, robertdeng@smu.edu.sg}
\vspace{-1em}}


\maketitle

\begin{abstract}
%

Cloud computing offers resource-constrained users big-volume data storage and energy-consuming complicated computation. However, owing to the lack of full trust in the cloud, the cloud users prefer privacy-preserving outsourced data computation with correctness verification. However, cryptography-based schemes introduce high computational costs to both the cloud and its users for verifiable computation with privacy preservation, which makes it difficult to support complicated computations in practice.

Intel Software Guard Extensions (SGX) as a trusted execution environment is widely researched in various fields (such as secure data analytics and computation), and is regarded as a promising way to achieve efficient outsourced data computation with privacy preservation over the cloud. But we find two types of threats towards the computation with SGX: Disarranging Data-Related Code threat and Output Tampering and Misrouting threat. In this paper, we depict these threats using formal methods and successfully conduct the two threats on the enclave program constructed by Rust SGX SDK to demonstrate their impacts on the correctness of computations over SGX enclaves. In order to provide countermeasures, we propose an efficient and secure scheme to resist the threats and realize verifiable computation for Intel SGX. We prove the security and show the efficiency and correctness of our proposed scheme through theoretic analysis and extensive experiments. Furthermore, we compare the performance of our scheme with that of some cryptography-based schemes to show its high efficiency.
\end{abstract}


\vspace{-0.5em}
\section{Introduction}
Public cloud computing service has become the first choice of more and more enterprises and users to deal with big data processing. However, security and privacy issues raise when the data are outsourced and handled over semi-trusted cloud. One crucial problem is that the processed results provided by the cloud platform may be incorrect. Verifiable computation is an essential way to address this concern, which studies how a prover in a semi-trusted cloud service provider (CSP) can convince a verifier that its computation is correct and not tampered\cite{yu2019verifiable}. But most existing schemes are constructed based on cryptography, which incur serious new issues: some schemes only support specific functions and cannot be used in general \cite{ma2015verifiable}; some generic schemes incur extremely high computation and communication costs, which are not applicable in practice for resource-constrained entities \cite{catalano2013practical}.

Another approach to guarantee the correctness of privacy-preserving computation is to use trusted execution environment (TEE) \cite{ekberg2013trusted} provided by trusted computing platform (TCP) \cite{smith2013trusted}. Recent advance in TCP such as Intel Software Guard Extensions (SGX) offers a TEE, called Enclave, which isolates code and data from untrusted regions within a device. An SGX-enabled processor protects the integrity and confidentiality of the computation inside an enclave by isolating the enclave’s code and data from privileged software and hardware in an outside environment, including the underlying operating system and hypervisor. With the security properties of the enclave signature and memory access control mechanism provided by the Intel SGX, all the codes in the enclave are considered untamed, not only the static code block before enclave initialization, but also the actual execution process. 

Compared with other TCP technology, Intel SGX provides a smaller Trust Computing Base (TCB), which makes its attack surface smaller than others. Hence, it becomes very popular in various institutes and industrial fields (such as Alibaba Cloud \cite{FortanixAlibabaCLoud} and Microsoft Research \cite{ohrimenko2016oblivious}). However, we find out two categories of threats towards SGX programs, i.e., Disarranging Data-Related Code (DDRC) threat, Output Tampering and Misrouting (OTM) threat. DDRC threat disarranges data-related code in order to make final computing results wrong. OTM threat tampers or misroutes the output of an enclave to a wrong next enclave in order to make the final computing result incorrect. Both the threats are owing to the fact that Intel SGX itself cannot guarantee the invocation order of the enclave functions that is determined outside the enclave not to be untamed. Moreover, if the output of an Enclave ECall function is returned to an untrusted domain, the output can be tampered by the attacker and then input to a next Enclave, which finally leads to a wrong result. To the best of our knowledge, there exist no solutions to defend these two threats to ensure accurate computation over Intel SGX. 

In this paper, we propose a scheme based on hash chain to record the topology of the Enclave ECall function execution order. Therefore, it can distinguish different execution orders of an execution plan. Hence our scheme can detect the DDRC threat. Furthermore, we combine the output result of an enclave to the hash chain to resist the OTM threat. If the output result coming into the next enclave is not the corresponding result, the hash chain will be changed and lead to the detection of the OTM threat. 

Generally speaking, our scheme offers an efficient verifiable computation mechanism for Intel SGX based privacy-preserving data computation by guaranteeing the invocation order of enclave functions. Our scheme can guarantee the basic requirement of data privacy in the scenario of computation-weak users delegating computation work to the computation-powerful cloud, and provides the user with an efficient way to verify the correctness of outsourced computation. Moreover, our scheme can support any types of data processing, operations and computations and greatly outperforms many conventional verifiable computation schemes based on cryptography. It can be commonly applied into many cloud computing services, such as batch computing services (Hadoop \cite{shvachko2010hadoop}), stream computing services (Storm, Spark, Flink \cite{carbone2015apache}), data processing and computation services and Database services (MySQL, Oracle) by implementing the services with a number of enclave functions, possibly run by multiple enclaves. Specifically, our contributions can be summarized as follows:

$\bullet$ We find two valid threats (DDRC threat and OTM threat) on the data computation over Intel SGX, which can affect the correctness of outsourced computation provided by the cloud equipped with SGX.

$\bullet$ We conduct experiments to attack the program constructed by Rust SGX SDK, and show the validity of the two threats on the current SGX-supported computation systems.

$\bullet$ We propose an efficient and secure scheme to perform verifiable computation for Intel SGX, which allow users to publically verify the computation results and thus successfully resist both DDRC threat and OTM threat.

$\bullet$ We compare our verifiable computation scheme with traditional cryptography-based verifiable computation schemes through experiments to further show its high efficiency.


\textbf{Roadmap}. The rest of this paper is organized as follows. Section \ref{sec2} briefly reviews related work, followed by preliminaries in Section \ref{sec3}. The proposed verifiable computation scheme for Intel SGX based privacy-preserving data computation is presented in Section \ref{sec4}. Section \ref{sec5} analyzes the security properties and computational complexity of our scheme. In Section \ref{sec6}, we conduct threats towards the enclave program constructed by Rust SGX SDK, and evaluate the effectiveness of our scheme through extensive experiments and comprehensive comparison. Finally, the last section concludes the whole paper.

\section{Related Work} \label{sec2}
In order to achieve the verifiability on the correctness of outsourced computation \cite{yu2017survey}, researchers strive to break through in two directions. One is verifiable computation based on cryptography, which often relies on mathematical and cryptographic techniques. The other is TEE-based scheme which relies on trusted execution environment and the security properties provided by TCP. 
\subsection{Verifiable Computation Based on Cryptography} 
The verifiable computation based on cryptography incurs high computational cost, which may influence the flexibility of functionality. Hence, two kinds of schemes were proposed and designed to balance between efficiency and functionality. The conventional verifiable computation cannot simultaneously satisfy both efficiency and flexibility.

\subsubsection{}Some solutions (i.e., \cite{yan2017context, yu2019verifiable, parno2013pinocchio}) aim to achieve generality by depending on Probabilistically Checkable Proofs (PCPs) or Fully Homomorphic Encryption (FHE). PCP is theoretically perfect, but it is difficult to be applied in practice for its high computation and memory overhead. Although there are many improved PCP-based schemes \cite{cormode2012practical, setty2012making}, they still have some drawbacks (e.g., high prover’s workload and heavy network overhead) with regard to performance. By combining the cryptography method and FHE, researchers have created many optimized verification schemes. Gennaro et al.\cite{gennaro2010non} first formalized the notion of verifiable computation and presented non-interactive verifiable computing by integrating Yao's Garbled Circuits with a FHE scheme. Yan et al. \cite{yan2017context} proposed a context-aware verifiable computing scheme that combines FHE and an auditing protocol to verify the computation result. Yu et al. \cite{yu2019verifiable} proposed a scheme which combined FHE and a polynomial factorization algorithm to allow public verification and provide privacy preservation on outsourced computation. However, these works suffer from high computation and memory overhead due to the low efficiency of FHE.

\subsubsection{}Some solutions are comparatively efficient but function-specific \cite{ma2015verifiable}. Kate et al. \cite{kate2010constant} and Benabbas et al. \cite{benabbas2011verifiable} proposed efficient constructions, but both only permit limited functions, i.e., polynomials and set operations, which are far away from the practical and flexible needs of the real-world scenarios. Hence, there are many non-interactive schemes \cite{gennaro2013fully, boneh2011homomorphic} that focus on supporting general functions, but there is still a long way to go.
%
\vspace{-0.5em}
\subsection{TEE-Based Solutions} 
The other research branch on verifiable computation is TEE-based solutions that depend on the security properties provided by the TEE. Duarte et al. \cite{duarte2018leveraging} presented SafeChecker, which is a mobile-oriented computation verification system and combines verifiable computation and ARM TrustZone together. SafeChecker uses ARM TrustZone as a secure storage container to store a private intermediate value during the execution of a verification system. Verification still uses SNARKs \cite{ben2013snarks} to generate an execution proof, which is the main reason of performance reduction. 
Flicker \cite{mccune2008flicker} allows verifiable execution, but there exist some obstacles when being deployed with specific TEEs. Schuster et al. \cite{schuster2015vc3} first proposed a verifiable computation scheme based on Intel SGX for MapReduce jobs in the semi-trusted cloud. But this VC3 scheme focuses on the data security and the integrity of the data transmission between mapper-nodes and reducer-nodes equipped with SGX-enabled processors, while ours concentrates on the integrity and security of the program executed in an SGX-enabled processor, especially the invocation order of the ECall functions and the data transmitted between enclaves. VC3 ignores the DDRC and OTM threats, which could affect the correctness of outsourced computation. Ryoan \cite{hunt2016ryoan}, as a distributed sandbox, allows users to process security data with untrusted software, but it does not solve DDRC and OTM threats systematically.

\section{Preliminaries} \label{sec3}
In this section, we first introduce the basic concepts of Intel SGX. Then, we describe the system model of our work, potential threats of using SGX for complicated computation at the cloud and our design goals.
\subsection{Basic Concepts}
Here we only illustrate necessary information related to our scheme and give several basic concepts used in the following scheme description.

\textbf{Local Attestation:} It occurs when two enclaves located in the same platform authenticate with each other. When an enclave needs to prove its identity to a target enclave, it uses \texttt{EREPORT} instruction to generate a REPORT structure according to its measurement-based and certificate-based identity (MRENCLAVE, MRSIGNER, ISVPRODID and ISVSVN), together with the message that it wants to deliver. Then the current enclave uses a symmetric key, which can only be obtained by the enclave, to compute the MAC tag of the REPORT structure. The target enclave will verify the MAC tag using the report key returned by \texttt{EGETKEY} instruction. Hence, the enclave proves its identity to the target enclave.

\textbf{Remote Attestation:} It occurs when an enclave gains the trust of a remote provider.
With Remote Attestation, a combination of Intel SGX software and platform hardware is used to generate a quote that is sent to a third-party server to establish trust. The software includes the application’s enclave, a Quoting Enclave (QE) and a Provisioning Enclave (PvE). Both of QE and PvE are provided by Intel. A digest of the software information is combined with a platform-unique asymmetric key from the hardware to generate a quote, which is sent to a remote server over an authenticated channel. If the remote server determines that the enclave is properly instantiated and is running on a genuine Intel SGX-capable processor, it can now trust the enclave and provision secrets over the authenticated channel.

\textbf{Enclave Signature and Measurement:} Intel Software Guard Extensions provides a signature mechanism, which takes place during the enclave building process. After the enclave is built, the enclave will have a signature, which is used to verify the integrity of the enclave itself before the actual enclave is executed, and the verification process is automatically executed by Intel SGX. If the verification fails, the initialization of the enclave will fail, which will cause the enclave program to fail to execute properly. This mechanism ensures that all enclaves that can be executed are not tampered. The security of this mechanism depends on the confidentiality of the signature key.

\subsection{System Model}\label{sysmodel}

\begin{figure}
	\centering
	\includegraphics[width=3.2in]{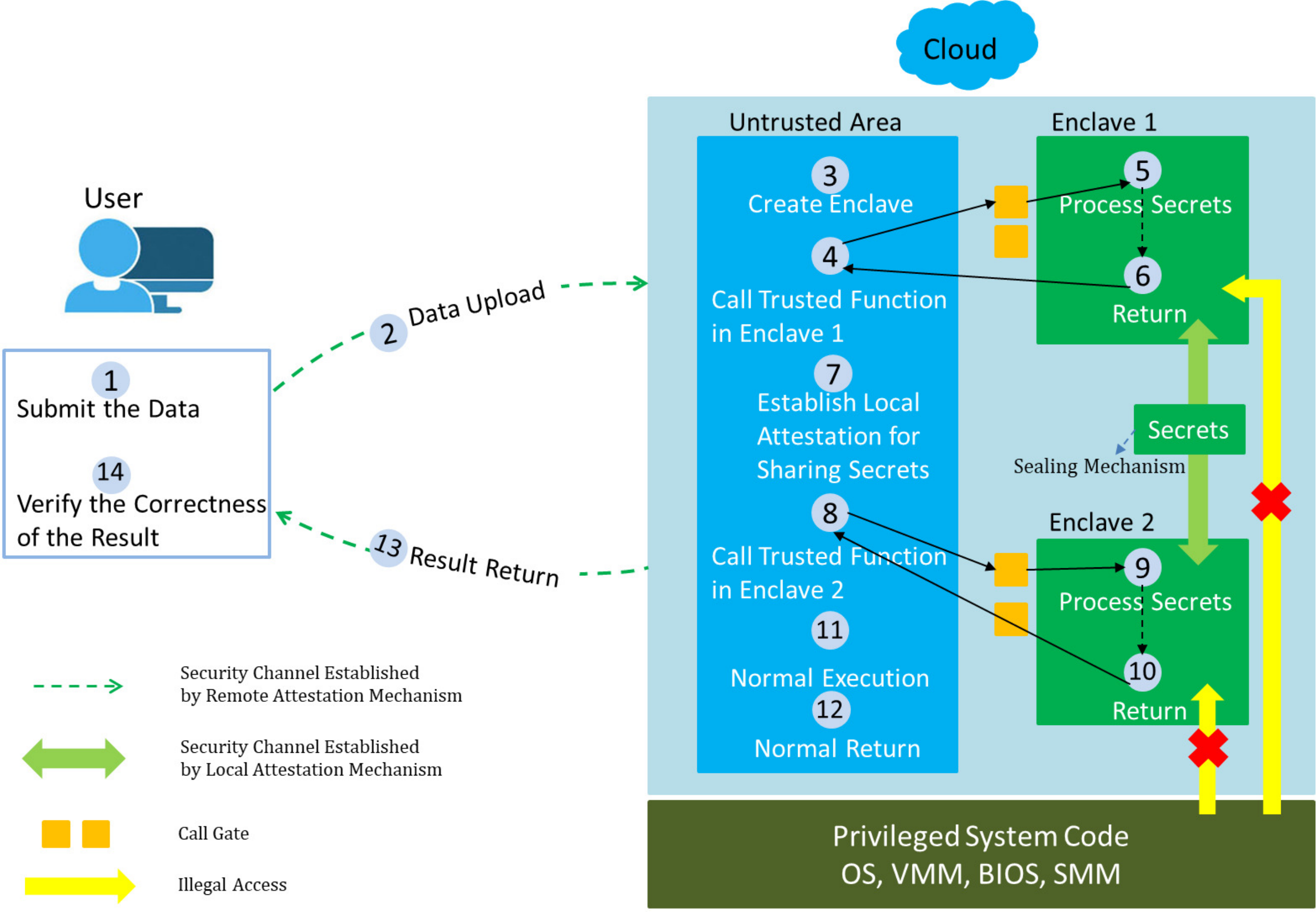}
	\caption{System Model}
	\label{figure1}
	\vspace{-2em}
\end{figure}

The system is composed of two types of entities as shown in Fig. \ref{figure1}:

\textbf{User:} The user submits a computation request to the cloud and also owns the ability to generate an execution plan based on a computation request and to verify the correctness of the result provided by the cloud. 

\textbf{Cloud:} The cloud is an SGX-enabled platform, which is responsible for processing the computation request issued by the user. But it could be a malicious entity, which means that it may be curious to obtain the private information of users, change intermediate results of the computation execution plan, and even interfere with or disarrange the normal invocation of Enclave ECall functions, e.g., for saving resources or exposing computation results.

To make a balance between the big volume of data and the limited size of enclave memory, multi-enclave setting is always employed. In our system, we assume that all the user computation request $Request$ can be resolved into basic functions (i.e., the generation process of an execution plan), and all the basic functions are implemented by Enclave ECall functions.  There exist many ad-hoc tools to generate an execution plan for a specific request (e.g., a SQL parser\cite{sqlparser}), the details of which will not be illustrated in this paper. In our system, we assume that the ECalls belong to different enclaves to achieve high flexibility. Herein, we denote difference enclaves in a cloud service scenario as a set $\{e_1,e_2,e_3,\cdots\}$, and $e_i$ represents a specific enclave in this scenario. The set $\{e^{1}_{i}, e^{2}_{i},\cdots\}$ represents different ECalls in enclave $e_i$ such that $e^{1}_{i}$ means the $1^{st}$ ECall function of the $i^{st}$ enclave with semantics $[\![e^{1}_{i}]\!]$. For a computation request $Request$, it can be resolved into a sequence of ECall functions $\rho=\langle e^{i'}_{i},\cdots,e^{j'}_{j},\cdots\rangle$, which we call execution plan in this paper. Hence, the semantic of $\rho$ is  $[\![\rho]\!]=\langle[\![e^{i'}_{i}]\!],\cdots,[\![e^{j'}_{j}]\!],\cdots\rangle$.

For a better understanding, we give a brief execution procedure between two enclaves. Fig. \ref{figure1} shows a normal execution example of one request about data processing and computation at the cloud. First, the user submits its data to the cloud. Then, the data is transmitted to the cloud through a secure channel established by the Remote Attestation (RA) mechanism provided by the Intel SGX. The cloud creates Enclaves needed for fulfilling the computation requested by the user, i.e., the generation process of execution plan. The details of the process at the cloud are depicted in the figure. After the execution, the processing result is transmitted to the user through the secure channel established by RA mechanism. The user verifies the correctness of the result and decides whether to accept the result according to verification.

\subsection{Potential Threats} \label{sec3-2}

%
%
%
%

\subsubsection{DDRC threat}
\begin{figure}
	\centering
	\includegraphics[width=2.2in]{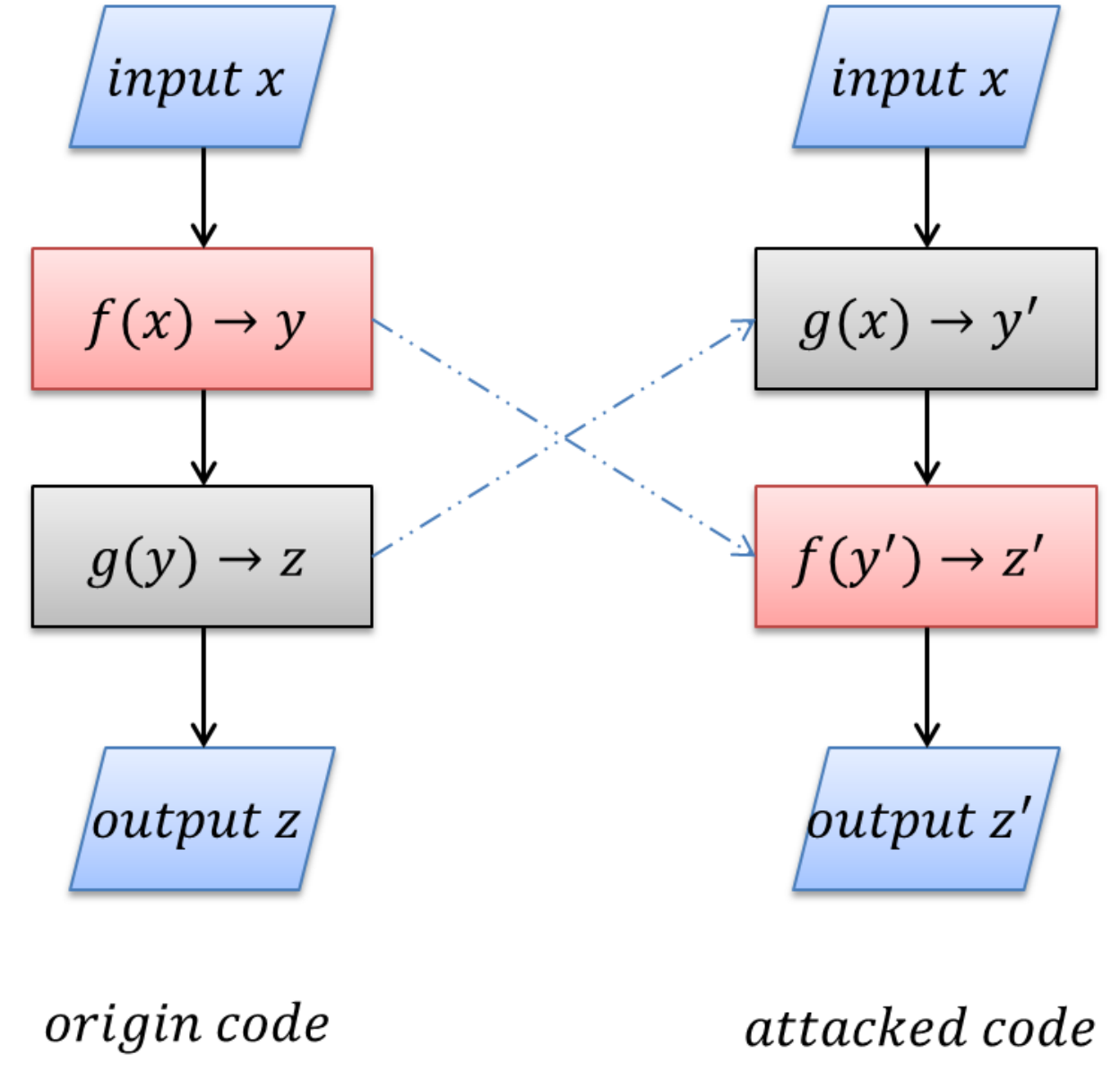}
	\caption{An example of DDRC threat}
	\label{figure2}
	\vspace{-1em}
\end{figure}

We show a DDRC threat towards Intel SGX programs. Fig. \ref{figure2} depicts an example of a DDRC threat. In the figure, the left part is origin code and the right part is attacked code. In this paper, we use $f(x)$ and $g(x)$ to represent Enclave ECall functions, which means that the execution of $f(x)$ and $g(x)$ is in enclave, the trusted part. Formally, $f(x)$ and $g(x)$ are denoted as $e^1_1$ and $e^1_2$ respectively, which is consistent with Fig. \ref{figure1}. Hence, the execution plan is $\rho=\langle e^{1}_{1},e^{1}_{2}\rangle$ with $[\![\rho]\!]=\langle[\![e^{1}_{1}]\!],[\![e^{1}_{2}]\!]\rangle$. As we can see from Fig. \ref{figure2}, the attacker disarranges the invocation order of $f(x) \rightarrow y $ and $g(y) \rightarrow z $ into $g(x) \rightarrow y^{\prime}$ and $f(y^{\prime}) \rightarrow z^{\prime}$. The actual execution plan is changed into $\rho'=\langle e^{1}_{2},e^{1}_{1}\rangle$ with $[\![\rho']\!]=\langle[\![e^{1}_{2}]\!],[\![e^{1}_{1}]\!]\rangle$. Obviously, $\rho\neq\rho'$ and $[\![\rho]\!]\neq[\![\rho']\!]$. Due to Intel SGX does not verify the integrity of the untrusted code, whereas only verifies the enclave code, the enclave signature mechanism of Intel SGX cannot find the disarrangement of the code. That is to say, SGX signature mechanism only verify the integrity $e_1$ and $e_2$, not the integrity of execution plan $\rho$. All in all, the execution is attacked. We illustrate the practicability of DDRC threat in Section \ref{attacksim}-a.


\subsubsection{OTM threat}
Output Tampering and Misrouting (OTM) Threat is the threat targeting on MapReduce platforms like Hadoop\cite{dinh2015m2r}. We found that it can also be conducted on Intel SGX programs towards the output of an enclave, which can influence the correctness of computations. We illustrate the practicability of OTM threat in Section \ref{attacksim}-b.

\textit{Output Tampering}: When the output result of an Enclave function is transmitted to the untrusted area, an attacker may duplicate, eliminate and tamper with the output result. As can be seen in Fig. \ref{figure1}, a malicious cloud app calls Enclave 2 and inputs wrong data output by Enclave 1. It is easy for the malicious cloud app to do this by just tampering with the output of Enclave 1.

\textit{Output Misrouting}: Also take Fig. \ref{figure1} as an illustration, a malicious cloud app can input the output of other Enclave that is not Enclave 1 into Enclave 2.

\subsection{Design Goal}

To defend the security threats mentioned in Section \ref{sec3-2}, we design a scheme that can check both the invocation order of an execution plan and data routing between enclaves. Side channel attack is out of the scope of this paper. In case the cloud implements replay attacks, we formulate our design goal as bellow:

1) The invocation order of an execution plan and data routing information should be checked to guarantee that the cloud executes the computation correctly;

2) Besides the invocation order, the scheme should also be able to guarantee the correct input data of enclaves in multiple enclave cooperation. It needs to distinguish the different inputs and requests. That is to say, the same request with different data and the same data with different requests should result in different evidence, e.g., hash chains;

3) The user can verify the correctness of hash chain from the cloud to perform cloud computation auditing. Specifically, the user can compute the hash without executing the operations that are executed in the cloud.
	

\vspace{-0.5em}
\section{Verifiable Computation via SGX} \label{sec4}
\vspace{-0.5em}
In order to resist the aforementioned threats in Section \ref{sec3-2}, we propose a secure verification scheme and discuss its executions in the scenarios of one and multiple enclaves. In this section, we first give an overview of our proposed scheme in Section \ref{sec4-A}. Then, we specifically describe our designed algorithms in Section \ref{sec4-B}.

\subsection{Proposed Scheme}\label{sec4-A}

Fig. \ref{figure4} gives a schematic overview of our verification scheme. Corresponding to the procedure in Fig. \ref{figure1}, herein we present the details of our scheme to defend the threats mentioned in Section \ref{sec3-2} as follows:

\begin{figure}
	\centering
	\includegraphics[width=3.5in]{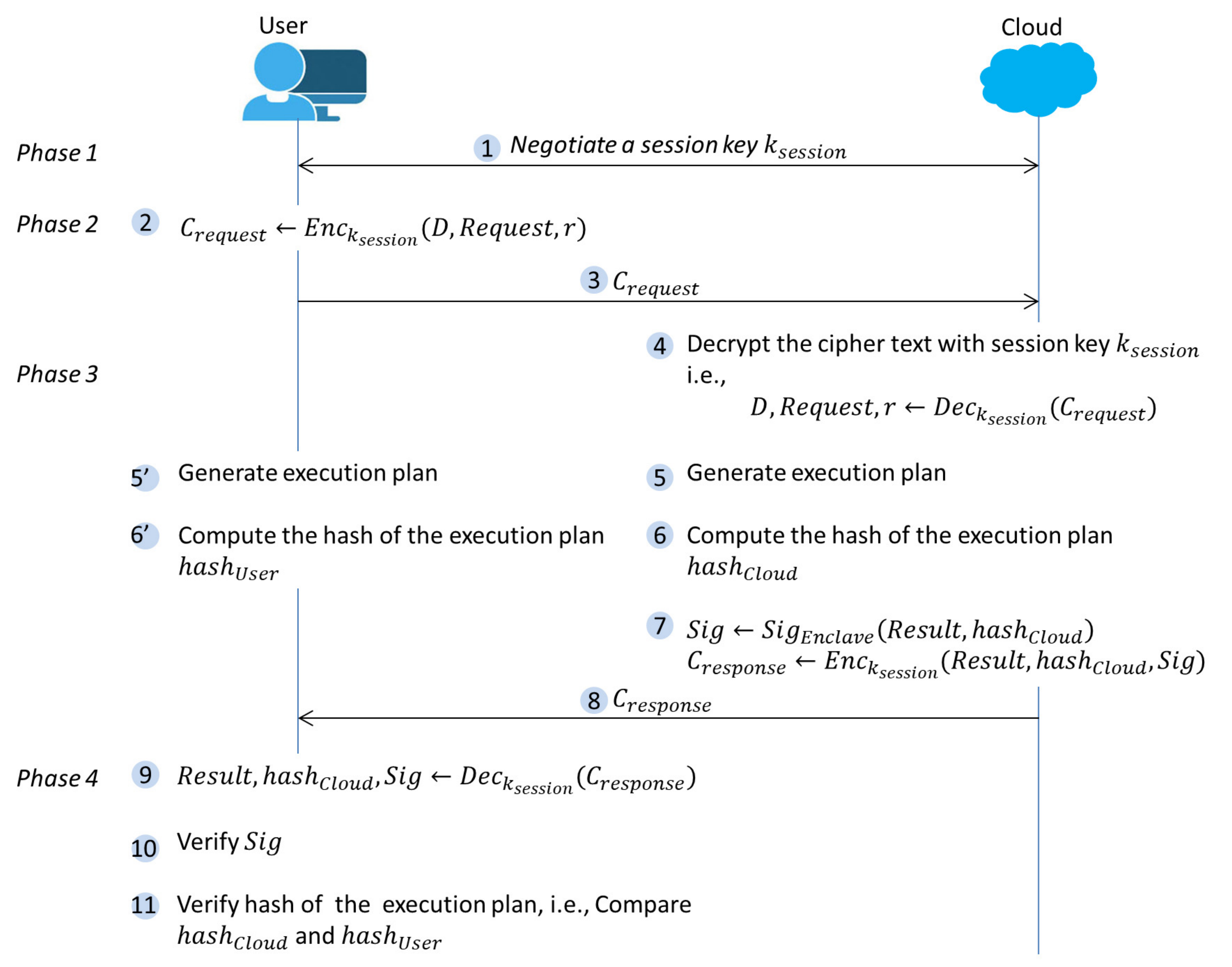}
	\caption{Proposed Verification Scheme}
	\label{figure4}
	\vspace{-1em}
\end{figure}

\subsubsection{System setup @All entities (Phase 1)}

This phase corresponds to step 1 in Fig. \ref{figure4}. The tags of Enclave ECall functions are public and initiated in system setup phase. the user issues a Remote Attestation with the cloud, which is supported by Remote Attestation mechanism of Intel SGX. After Remote Attestation, the user and the cloud share a master secret, which can be used to derive a session key $k_{session}$, and a signature key pair $sk_{Cloud}$ and $vk_{User}$.

\subsubsection{Request submitting phase @User (Phase 2)}
In the process of an operation request, the user first generates a random number $r$. Then it encrypts data $D$, $Request$ and $r$ with session key $k_{session}$ into ciphertext $C_{request}$. Then the user submits cipher $C_{request}$ to the cloud. This phase contains steps 2 and 3 in Fig. \ref{figure4}.

\subsubsection{Request process and result return phase @Cloud (Phase 3)}
At the cloud end, this phase contains steps from 4 to 8. After receiving $C_{request}$ from the user, the cloud loads $C_{request}$ into the enclave (i.e., $AttestEnclave$), and decrypts it with the session key $k_{session}$. Then it generates the corresponding execution plan to specify the selected enclaves according to $Request$. Data $D$ is transmitted from $AttestEnclave$ to the selected enclaves through Local Attestation. Next, the cloud executes two works in the enclaves: it processes the data $D$ according to the execution plan; and then it computes the hash of the execution plan according to the algorithm illustrated in Section \ref{algo-cloud}. After the whole execution, it obtains $Result$ and $hash_{cloud}$ and then delivers them to the $AttestEnclave$ through the secure channel established by Local Attestation. $AttestEnclave$ first signs $Result$ and $hash_{cloud}$ using $sk_{Cloud}$, and then encrypts them into $C_{response}$ with session key $k_{session}$. At last, $AttestEnclave$ sends the signed ciphertext $C_{response}$ to the user.

\subsubsection{Verification phase @User (Phase 4)}
This phase contains steps 5' and 6' and steps from 9 to 11 in Fig. \ref{figure4}. In steps 5' and 6', the user generates the corresponding execution plan of the current request, and computes the hash of the execution plan according to the algorithm illustrated in Section \ref{algo-user} to obtain $ hash_{User} $. In step 9, The user decrypts $ C_{response} $ transmitted from the cloud with session key $ k_{session} $ and obtains $ Result $, $ hash_{Cloud} $, $ Sig $. In step 10, the user verifies the validity of $ Sig $ using $vk_{User}$. It compares $ hash_{user} $ and $ hash_{Cloud} $ in step 11 to verify the correctness of the hash of execution plan. If they are equal, $result$ is correct, otherwise, $result$ is wrong. 

\begin{table}
	\caption{Notation Description}\label{tab1}
	\centering
	\begin{tabular}{|l|p{6cm}|}
		\hline
		\textbf{Symbols} & \textbf{Description}\\
		\hline
		$ \oplus $ & The xor operation; \\
		$\mathbf{H}$ &  The $ m $-dimensional vector listing the hash code of each predecessor node; \\
		$H_{i}$ &  The hash code of $ i $-th predecessor node; \\
		$\mathbf{res}$ &  The $ m $-dimensional vector listing the result of each predecessor node; \\
		$res_{i}$ & The result of $ i $-th predecessor node; \\		
		$res_{current}$ & The result of the current function; \\		
		$H_{current}$ & The hash code of the current function. \\		
		\hline
	\end{tabular}
	\vspace{-1em}
\end{table}

\subsection{Two Algorithms of Hash Computation}\label{sec4-B}
\subsubsection{How to compute the hash of the execution plan at the cloud}\label{algo-cloud}

Before going into detailed description, we first list the notations used in this section in Table \ref{tab1}. In this part, we introduce the algorithm used at the cloud in step 6 in Fig. \ref{figure4}. This algorithm is used to compute the hash code of each Enclave ECall implementation function within an execution plan. Specifically, when executed to the last function of the execution plan, the corresponding hash code can be used to identify the whole execution plan. Herein, we first introduce how an ECall function to calculate its hash in different cases, and then we present the algorithm in details.

\par\noindent (a) Hash computation in different cases


Besides the computing services under the expectation of the user, ECall function should use its function tag to record the invocation order, which lays the basis for verifiable computation. It first needs to combine its tag with the hash code from the previous function to record the topology of the Enclave ECall function execution order and generate its hash input $h_i$; and then it calculates its own hash code and forwards it to the next ECall function.

Herein, we first discuss how the current ECall function to record the execution order and generate hash input. We classify them into four cases according to the relationship between the current ECall function and its predecessor as shown in Fig.~\ref{figure5}. We use $H_i$ to indicate the hash code of ECall function $f_i$, which is introduced in the following part:

\begin{figure}
	\centering
	\includegraphics[width=3.4in]{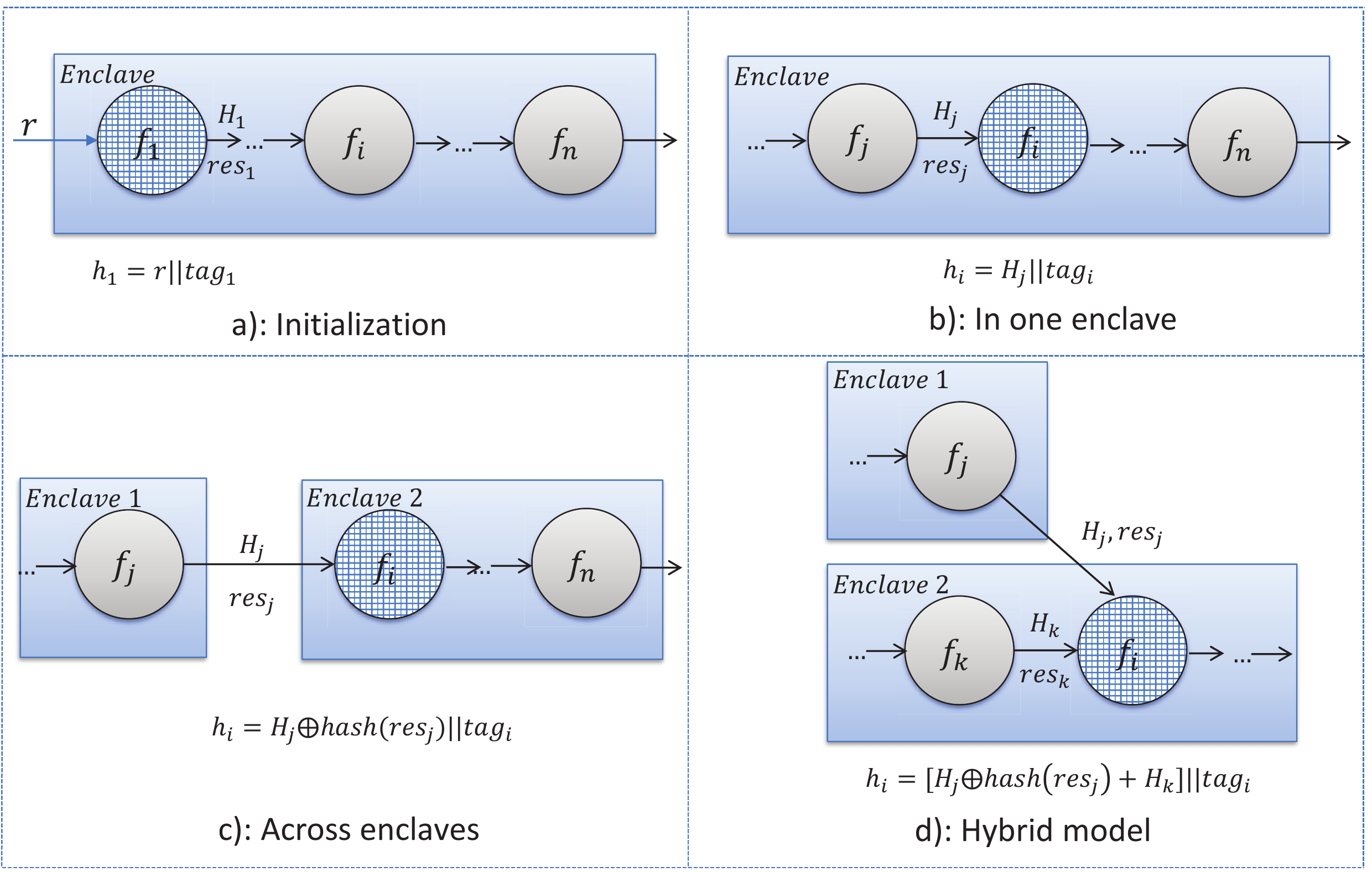}
	\caption{Hash input of current ECall function: a) Initialization; b) In one enclave; c) Across enclaves; d) A hybrid model}
	\label{figure5}
\end{figure}

\begin{figure}
	\centering
	\includegraphics[width=2.5in]{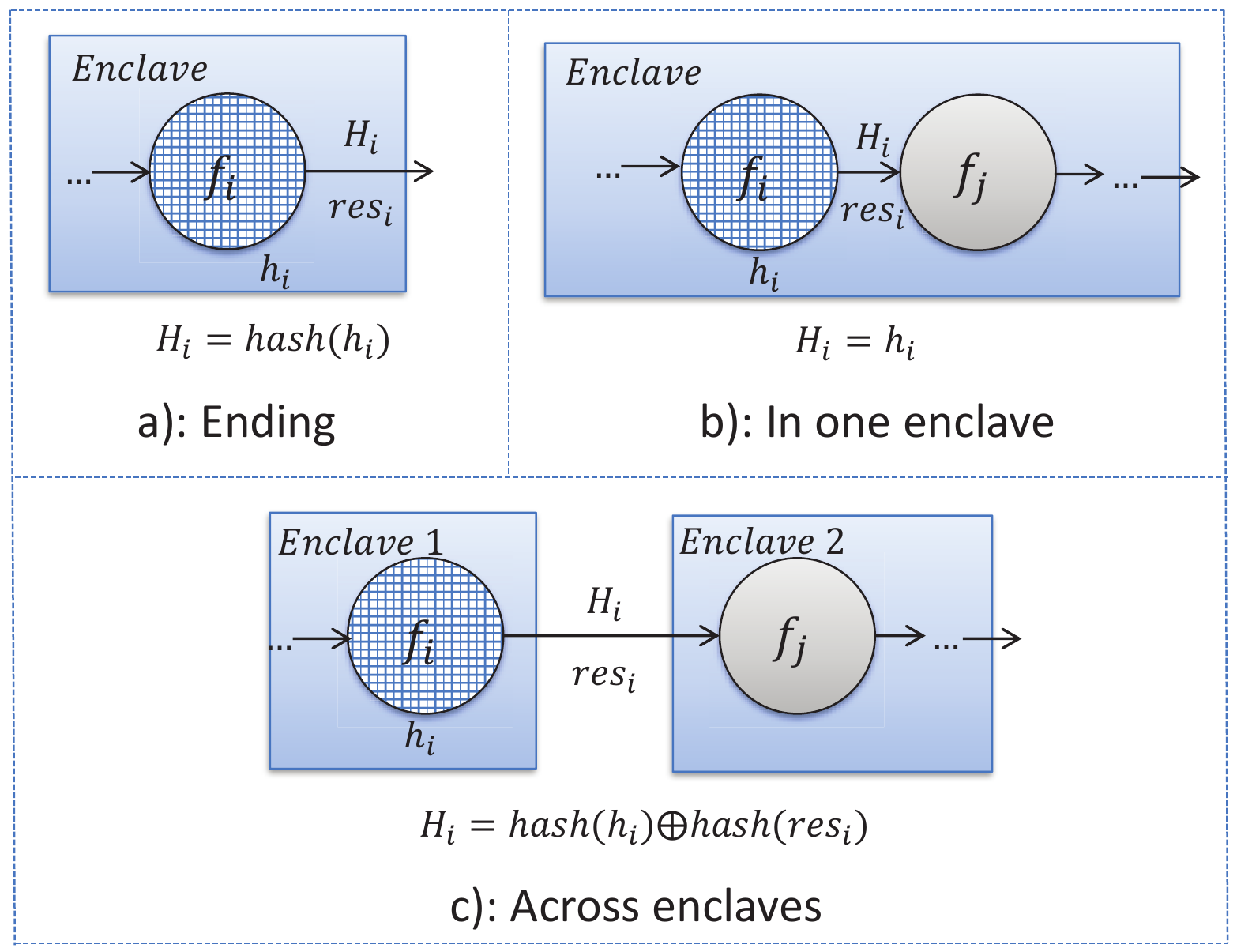}
	\caption{Hash input of current ECall function: a) Ending; b) In one enclave; c) Across enclaves}
	\label{figure6}
	\vspace{-1em}
\end{figure}

Case a): If the current ECall function is the first invoked function as shown in Fig.~\ref{figure5}a), only random number is input to initialize the computation, hence, ECall function can simply concatenate the random number and its function identifier to get its hash input $h_1=r||tag_1$.

Case b): If the current ECall function $f_i$ and its predecessor $f_j$ exist in the same enclave as shown in Fig.~\ref{figure5}b), its hash input also results in $h_i=H_{j}||tag_{i}$.

Case c): If its predecessor exists in another enclave as shown in Fig.~\ref{figure5}c), the current ECall function $f_i$ needs to execute the xor operation with the hash code of the computation result from previous function $f_j$. Therefore, its hash input becomes $h_i=H_j \oplus hash(res_j)||tag_{i}$.
	
Case d): As is shown in Fig.~\ref{figure5}d), the current ECall function receives output results from multiple functions. In this case, it needs to combine all results to get one hash input. Similar to the computation in the three cases above, the ECall function $f_i$ can get its hash input as $h_i=[H_j \oplus hash(res_j) +H_k] ||tag_i$.

Next, we elaborate on how to get the hash result of ECall function $f_i$ denoted by $H_i$. Different from the four cases above, the hash code computations only have three different cases as shown in Fig.~\ref{figure6}. After getting the hash input of each ECall function, how to gain the hash code of each function depends on its relationship with its following ECall function:

Case a): If the current ECall function is the last one for an execution plan, it simply calculates its hash result with the hash input $h_i$ and obtains $H_i=hash(h_i)$.

Case b): If the ECall function is the last one and its following ECall function is in the same enclave, then it does not need to do hash computation but sets hash code as its hash input $H_i=h_i$.

Case c): If its following ECall function is in another enclave, its hash code involves in the hash code of output result and results in $H_i=hash(h_i) \oplus hash(res_i)$.

Specifically, when executed to the last function of the execution plan, the corresponding output can be used to identify the whole execution plan.
\vspace{0.5em}

\par \noindent (b) Algorithm 1: The hash computation executed at the cloud

\begin{figure}
	\centering
	\includegraphics[width=3.4in]{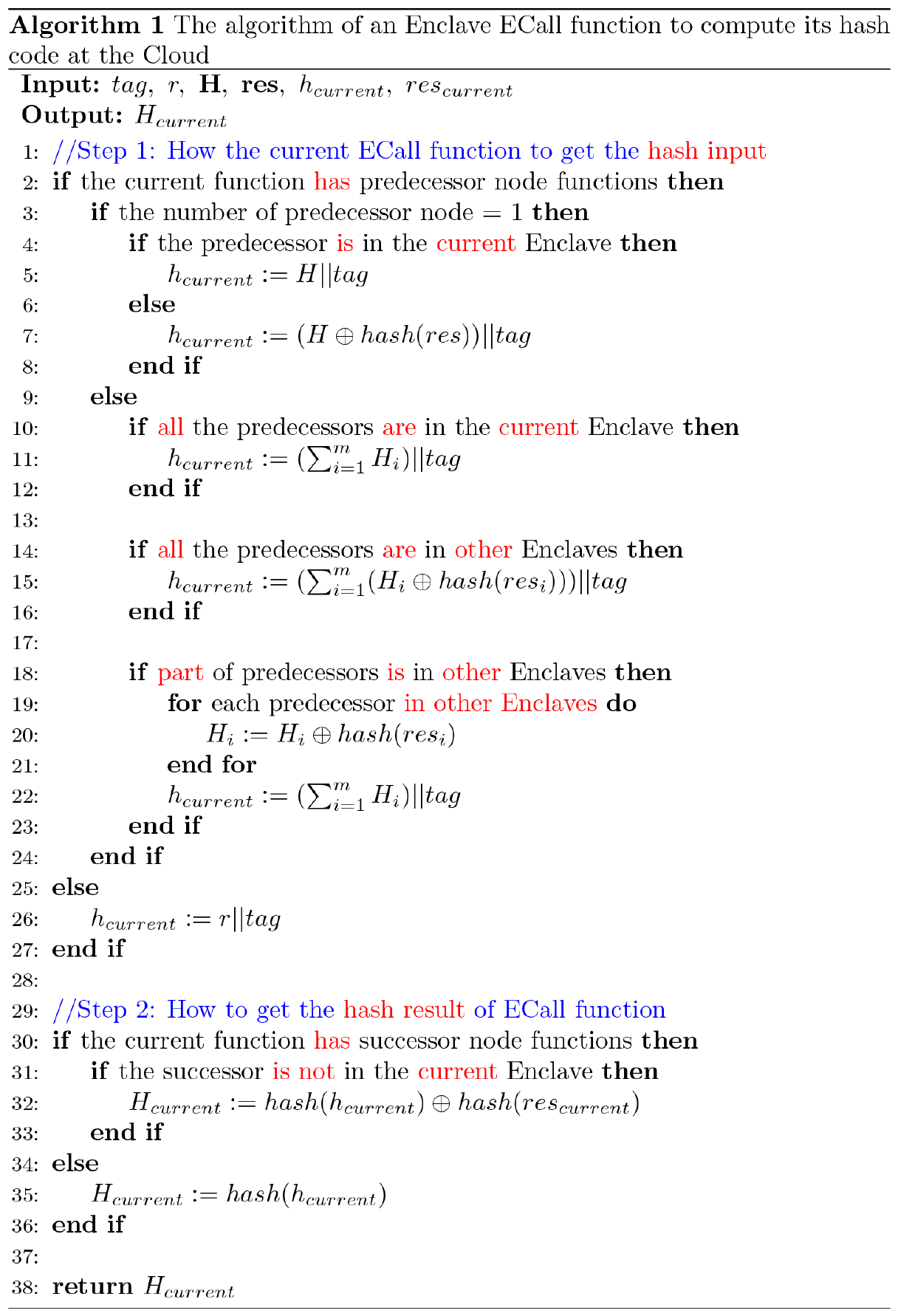}		
	\caption{The algorithm of an Enclave ECall function to compute its hash code executed at the cloud}
	\label{figure10}
	\vspace{-1em}
\end{figure}

We present Algorithm 1 for an Enclave ECall function to compute its hash code at the cloud in Fig. \ref{figure10}. Each Enclave ECall function executes this algorithm to obtain the corresponding hash code for each function in the entire execution plan.

\begin{figure}
	\centering
	\includegraphics[width=3.4in]{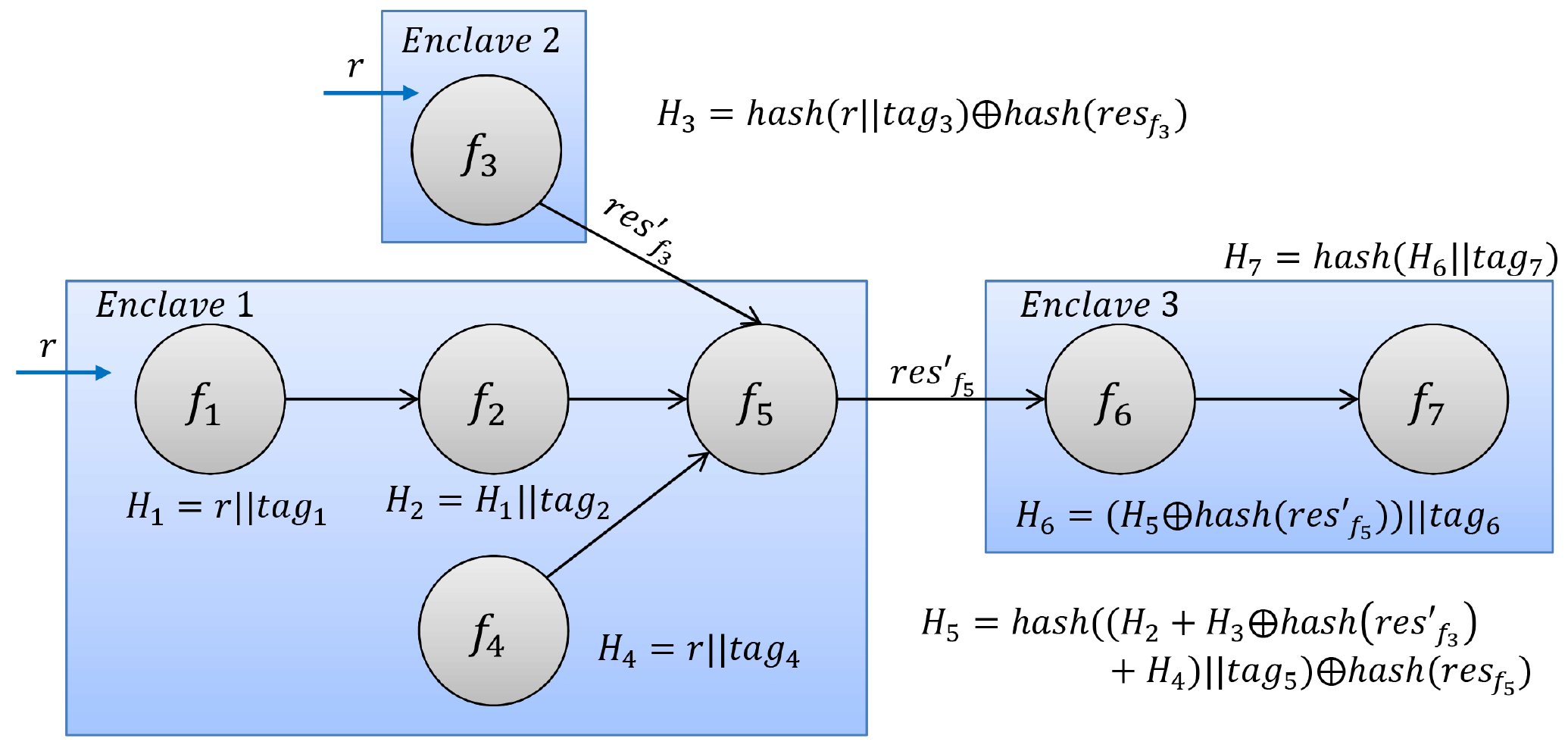}					
	\caption{A hybrid example}
	\label{figure9}
\end{figure}

Next, we present a hash code computation example in Fig. \ref{figure9}. The seven functions in hybrid example satisfy different cases in Figs. \ref{figure5} and \ref{figure6}, which is corresponding to several different situations in Algorithm 1. According to its corresponding situation, the hash code of each function is shown in Fig. \ref{figure9}. For this hybrid example, the cloud obtains the overall hash:
\begin{small}
	\begin{flalign*}
	\begin{split}
	hash_{cloud} 
	&= hash(((hash(((r||tag_1||tag_2) + ((hash(r||tag_3)\\
	&~~\oplus hash(res_{f_3}))\oplus hash(res_{f_3}^{'})) + (r||tag_4)) || \\ &~~tag_5)\oplus hash(res_{f_5}) \oplus hash(res_{f_5}^{'}) )||tag_6)||tag_7)\\
	\end{split}&
	\end{flalign*}
\end{small}

\begin{figure}
	\centering
	\includegraphics[width=3.4in]{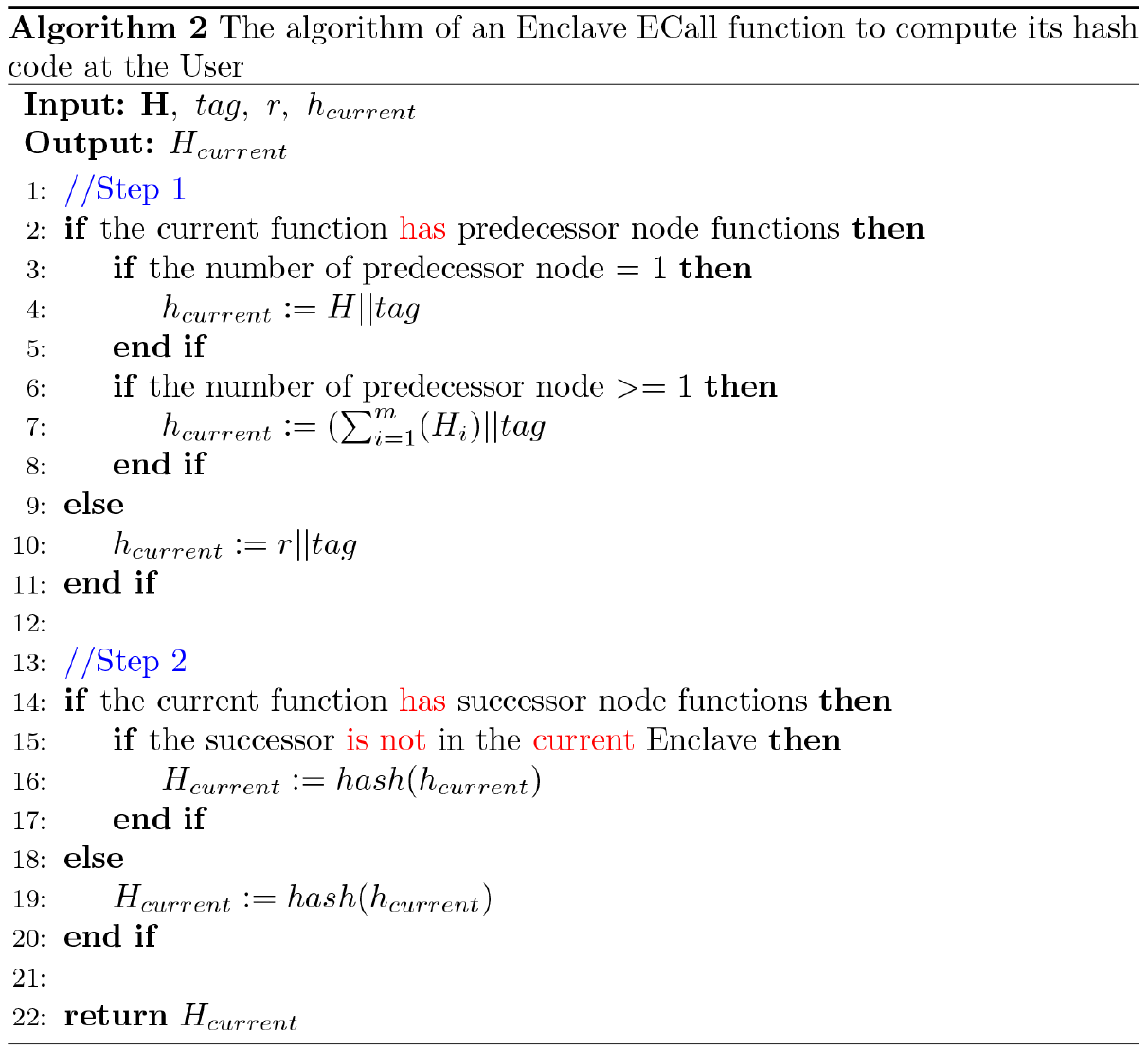}		
	\caption{The algorithm of an Enclave ECall function to compute its hash executed at the user}
	\label{figure11}
	\vspace{-1em}
\end{figure}


\subsubsection{How to compute the hash of the execution plan at the user}\label{algo-user}
In Fig. \ref{figure11}, we present Algorithm 2 to compute the hash code for each Enclave ECall function at the user.

There are two main steps in Algorithm 2 as we can see in lines 1 and 13 separately. In the algorithm, the predecessor node function is the one executed before the current function, and the condition in lines 14 and 15 means that after the execution of the current function, either the current function is the last function of this execution plan, or the next function to be invoked is not in the current enclave.

Next, we still take Fig. \ref{figure9} as an example. The hash code of each function is as follow: $H_1 = r||tag_1$, $H_2 = H_1||tag_2$, $ H_3 = hash(r||tag_3)$, $ H_4 = r||tag_4 $, $H_5 = hash((H_2 + H_3 + H_4) || tag_5)$, $H_6 = H_5||tag_6$, $H_7 = hash(H_6||tag_7)$.
For this hybrid example, the user obtains the overall hash:

\begin{small}
	\begin{flalign*}
	\begin{split}
	hash_{user}
	&= hash(((hash(((r||tag_1||tag_2) + (hash(r||tag_3))\\
	&~~~~+ (r||tag_4)) || tag_5))||tag_6)||tag_7)\\
	\end{split}&
	\end{flalign*}
\end{small}

Here, we illustrate the equivalence of $hash_{user}$ and $hash_{cloud}$. The xor operation has a special property. Formally, let $a$, $b$ be two numbers in $\mathbb{Z}$, $a\oplus b\oplus b=a$. In the hybrid example of Fig. \ref{figure9}, if $res_{f_{i}}$ is equal to $res'_{f_{i}}$ for each $f_{i}$, then $hash_{user}=hash_{cloud}$.

\section{Security Analysis and Computational Complexity}
\label{sec5}
In this section, we analyze the security of our design and show the computational complexity of Algorithms 1 and 2.


\subsection{Security Analysis}\label{sec5-1}
In this subsection, we mainly focus on theoretic analysis to show the advantages of our design, which can resist some popular attacks in existing systems.

\textbf{Resist replay attack.}
As discussed above, the computation inside the enclave can be guaranteed, while others outside the enclave may be tampered. The semi-trusted cloud may also execute a replay attack when it needs to complete the same computation task. But in our design, we choose different random numbers to initialize the computation of the hash chain. Only when the random numbers held by the user and the cloud keep consistent, will the verifiability check pass. The inserted random number can help resist the replay attack. 

\textbf{Correctness.}
In our system, we believe that the essence of verifiable computation is to authenticate a certain execution plan, i.e., identify a Directed Acyclic Graph. As we mentioned in Figs. \ref{figure5} and \ref{figure6}, we give all possible cases which the predecessors and successors of an Enclave ECall function belong to and conduct the corresponding calculations in two algorithms.  We consider all possible topological cases of the DAG in both two algorithms. In each case of Algorithm 1 at the cloud, we can find the corresponding case and computation part in Algorithm 2. We take all the required parameters of Algorithm 2 into consideration and transmit them from the cloud to the user. Therefore, our scheme can obtain the same hash of the specific execution plan at both the cloud and the user if the computation at the cloud follows the execution plan of the user. Hence, our system is correct. In Section \ref{sec6-B}, we conducted a corresponding experiment to verify the correctness of our scheme.
\subsection{Computational Complexity}
\vspace{-0.5em}
\begin{table}[h]
	\caption{Computation Symbol Description}\label{tab2}
	\centering
	\begin{tabular}{|l|p{5.5cm}|}
		\hline
		\textbf{Symbols} & \textbf{Description}\\
		\hline
		$ Hash $ &  The hash operation \\
		$ Xor $ & The xor operation \\
		$ Add $	& The addition operation \\
		$ Con $ & The concatenation operation \\
		$ Enc_{sgx} $ & The encryption operation at the cloud\\
		$ Dec_{sgx} $ & The decryption operation at the cloud\\
		$ Sig_{sgx} $ & The signature operation \\
		$ Ver_{sgx} $ & The verification operation \\
		$ LA_{sgx} $  & The Local Attestation operation \\
		$ RA_{sgx} $  & The Remote Attestation operation \\
		$ Enc $ & The encryption operation at the user\\
		$ Dec $ & The decryption operation at the user\\
		\hline
	\end{tabular}
\end{table}

We consider the operations listed in Table \ref{tab2} to analyze the computational complexity of our scheme shown in Fig. \ref{figure4}.

\par\noindent (a) Complexity of Algorithm 1

We consider an execution plan with $n$ Enclave ECall functions. For the situations that each function has more than one predecessors, we make the following assumption. Every function has at most $m$ predecessors, and there are at most $p$ predecessors not in the same enclave as the current function. Hence, the range of values for the two variables is $ 2 <= m <= n,~0 < p < m $. Every function has at most $q$ successors, and there are at most $q'$ successors not in the same enclave as the current function. For Algorithm 1, we can obtain that lines 11, 15 and 22 can be represented by $( m - 1 ) * Add + m' * ( Xor + Hash ) + 1 * Con $, and note that the range becomes $ 2 <= m <= n,~0 <= m' <= m $.

\begin{table}
	\caption{Computational Complexity of Algorithm 1}\label{tab4}
	\centering
	\begin{tabular}{|l|p{4.5cm}|l|}
		\hline
		\textbf{Line} & \textbf{Computational Complexity} & \textbf{Probability} \\
		\hline
		5  &  $ 1 * Con $ & $Pr_{1}$\\
		7  & $ 1 * ( Xor + Hash + Con ) $ & $Pr_{2}$ \\
		11/15/22  & $( m - 1 ) * Add + m' * ( Xor + Hash ) + 1 * Con $ & $Pr_{3}$\\
		26  & $ 1 * Con $ & $Pr_{4}$\\
		\hline
		32  & $ 2 * Hash + 1 * Xor $ & $(n-1)/n$\\
		35  & $ 1 * Hash $ & $1/n$\\
		\hline
	\end{tabular}
\end{table}

For Algorithm 1, the computational complexity in line 5, line 7 and line 26 is shown in Table \ref{tab4}. We can see that in step 1 of Algorithm 1, there are four possible categories of cases, which are lines 5, 7, 11/15/22, and 26. Suppose that the probability of each function node's category falling into four cases is $Pr_{1}$, $Pr_{2}$, $Pr_{3}$, $Pr_{4}$, and $Pr_{1}+Pr_{2}+Pr_{3}+Pr_{4}=1$ holds. For one function, the whole computation overhead $O^{Alg1}_{step1}$ in step 1 of Algorithm 1 is:
\begin{small}
$Con+Pr_{3}*(m-1)*Add+(Pr_{2}+Pr_{3}*m')*(Xor+Hash)$.
\end{small}

We assume that only one node could be the last node. Its computational costs in step 2 of Algorithm 1 is  $O^{Alg1}_{step2'} = 1*Hash$. For $n-1$ functions with successors, every successor not in the current enclaves needs a Local Attestation operation $LA_{sgx}$. Hence, for one function with successors, its computational complexity  $O^{Alg1}_{step2}$ in step 2 of Algorithm 1 is:
\begin{small}
$q'*(2*Hash+1*Xor+LA_{sgx})$.
\end{small}
For $n$ nodes, the whole computational cost $O^{Alg1}_{exe}$ of the execution plan is :
\begin{small}
	\begin{flalign*}
	\begin{split}
	O^{Alg1}_{exe}
	&=n*Con+n*Pr_{3}*(m-1)*Add\\
	&~~+(n*(Pr_{2}+Pr_{3}*m')+(n-1)*q')*Xor\\
	&~~+(n*(Pr_{2}+Pr_{3}*m')+1+2*(n-1)*q')*Hash\\
	&~~+(n-1)*q'*LA_{sgx}
	\end{split}&
	\end{flalign*}
\end{small}
The computation overhead at the cloud is $O_{Cloud} = O^{Alg1}_{exe} + 1*(Enc_{sgx}+Dec_{sgx}+Sig_{sgx}+RA_{sgx})$.
\vspace{0.5em}
\par\noindent (b) Complexity of Algorithm 2

The symbol set is the same as part (a). We list the computational complexity of Algorithm 2 in Table \ref{tab5}. 
\begin{table}[h]
	\caption{Computational Complexity of Algorithm 2}\label{tab5}
	\centering
	\begin{tabular}{|l|l|l|}
		\hline
		\textbf{Line} & \textbf{Computational Complexity} & \textbf{Probability}\\
		\hline
		4  &  $ 1 * Con $ & $Pr_{1}+Pr_{2}$\\
		7  &  $ (m-1)*Add+1 * Con $ & $Pr_{3}$\\
		10  &  $ 1 * Con $ & $Pr_{4}$\\
		\hline
		16 & $ 1*Hash $ & $(n-1)/n$\\
		19 & $ 1*Hash $ & $1/n$\\
		\hline
	\end{tabular}
\vspace{-0.5em}
\end{table}

The computational cost in steps 1 and 2 of Algorithm 2 and the whole computational cost of the execution plan $O^{Alg2}_{exe}$ are listed as below:
\begin{small}
	\begin{flalign*}
	\begin{split}
	O^{Alg2}_{step1}
	&=1*Con+Pr_{3}*(m-1)*Add\\
	O^{Alg2}_{step2}&=q'*1*Hash~~~~~~~~~~~~~~~~~O^{Alg2}_{step2'}=1*Hash\\
	O^{Alg2}_{exe}
	&=n*Con+n*Pr_{3}*(m-1)*Add\\
	&~~+(1+(n-1)*q')*Hash
	\end{split}&
	\end{flalign*}
\end{small}
The computation overhead at the user is: $O_{User}=O^{Alg2}_{exe}+1*(Ver_{sgx}+RA_{sgx}+Enc+Dec)$.

\section{Attack Simulation and Performance Evaluation}\label{sec6}

In this section, we prove the existence of the threats in Section \ref{sec3-2} through extensive experiments, and then show the correctness and capability of the our scheme. Moreover, we compare our scheme with a traditional verifiable computation scheme based on homomorphic encryption to show its high efficiency. 

The experiments were performed on a 64-bit, Intel 6-Core I5-8500 CPU clocking at 3.00GHZ with 8 GB RAM and running Ubuntu-16.04 and Docker-17.05.0-ce. For the SGX environment, we used Rust SGX SDK v1.0.8 in a docker image named \texttt{sgx-rust} provided by BaiduXLab\cite{github.com}. 

\subsection{Attack Simulation}\label{attacksim}

In this part, we present DDRC and OTM threat simulation on a SGX-based program constructed by Rust SGX SDK\cite{github.com}.
%
%
We identified that the application constructed by Rust SGX SDK is still vulnerable in face of DDRC and OTM threat.

\par\noindent (a) DDRC Threat Simulation

%
%

We used the hello-rust project provided by the Rust SGX SDK as the attack target. 
In order to maintain the original security properties of the project and show the effect of the attack more clearly, we made the following three minor changes on the original project: 
\subsubsection{}For the Enclave.edl file, we added an ECall function named \texttt{say\_something\_b()}.
and renamed \texttt{say\_something()} to \texttt{say\_something\_a()}. The arguments of \texttt{say\_something\_b()} are consistent with those of \texttt{say\_something\_a()}.
\subsubsection{}For the lib.rs file, we renamed \texttt{say\_something()} to \texttt{say\_something\_a()} and added the implementation of \texttt{say\_something\_b()}, whose function body was copied from that of \texttt{say\_something\_a()}. The function arguments are consistent with \texttt{say\_something\_a()}.
We added the corresponding flag in the output part of \texttt{say\_something\_b()} to indicate that it is executing.
\subsubsection{}For the main.rs file, we renamed \texttt{say\_something()} to \texttt{say\_something\_a()} and copied the invocation part of the \texttt{say\_something\_a()} in the main function and pasted it after the invocation part of \texttt{say\_something\_a()}. Then we renamed the pasted copy to \texttt{say\_something\_b()}, as the invocation to \texttt{say\_something\_b()}.

Up to now, all the modification to the source code has been done. We name it as the modified project in the following content. After compiling, Rust SGX SDK constructed a binary file called $app$. The output of $app$ was the successive output of \texttt{say\_something\_a()} and \texttt{say\_something\_b()}. We then conducted the binary analysis on $app$ by using two Unix disassembly tools, \textbf{\texttt{objdump}} and \textbf{\texttt{readelf}}, to analyze $app$ in the shell. The tool \textbf{\texttt{readelf}} can output information about the specified ELF file (executable and linkable format, the executable file format under the Linux platform). We used the following shell command, \texttt{readelf -a app > elf.txt}, 
to generate the elf information for the $app$ and observed the elf.txt, which showed that $app$ contained 256 entries, located in $ symbol~table~'.dynsym' $. We can find \texttt{say\_something\_a} in $ symbol~table~'.dynsym' $, and then located the entry address of the function name \texttt{say\_something\_a}, which was \texttt{0x00af90}. Similarly, we can get the entry address of \texttt{say\_something\_b()}, which was \texttt{0x00aff0}. Next, we can find the offset value corresponding to the entry address \texttt{0x00af90} and \texttt{0x00aff0} in $ relocation~section~'.rela.dyn' $ of the elf.txt file. The entry address \texttt{0x00af90 } and \texttt{0x00aff0} corresponded to the offset value \texttt{0x23edd8} and \texttt{0x23eee8} respectively. Based on these two addresses, we can locate the binary code corresponding to the location of the two ECall function invocations in the $app$. Then we used the following shell command, \texttt{objdump -d app > dump.txt}, 
to disassemble the binary $app$ and observed the dump.txt file, which contained five sections, i.e., $.init~.plt~.plt.got$ $.text~.fini$. The binary codes corresponding to \texttt{0x23edd8} and \texttt{0x23eee8} can be observed in $ section~.text $ respectively: \texttt{9952: ff 15 80 54 23 00}, \texttt{9958: 89 84 24 90 00 00 00}, \texttt{9aac: ff 15 36 54 23 00}, \texttt{9ab2: 89 44 24 40}.

In the file dump.txt, \texttt{9952} and \texttt{9aac} are the value of a kind of Special Function Registers named $\%rip$ in AT\&T assembly grammar. Hence, the value of $\%rip$ of the next assembly instructions are \texttt{9958} and \texttt{9ab2} respectively. \texttt{ff15} is the corresponding binary code of assembly instruction $jump$. The rest is the offset address. Due to Linux little endian mode, the following hexadecimal addition is true: \texttt{9958 + 235480 = 23edd8, 9ab2 + 235436 = 23eee8}.
Note that this hexadecimal addition skips the overflow. The tampering of the ECall function invocation sequence can be implemented by modifying the offset after \texttt{ff15} to an appropriate value. In this example, we used \textbf{\texttt{vim}} to modify the content of the binary file $app$. To be specific, we modified the corresponding offset values respectively from \texttt{ff1580542300} to \texttt{ff1590552300}, and from \texttt{ff1536542300} to \texttt{ff1526532300}. Here we denote the modified binary file as $app'$. After the modification, we used \textbf{\texttt{objdump}} to analyze $app'$ and we can find that the corresponding two lines has changed: \texttt{9952: ff 15 90 55 23 00}, \texttt{9aac: ff 15 26 53 23 00}.

Up to now, $app$ was attacked. 
Before attacked, the program $app$ first outputs the result of $ \texttt{say\_something\_a()} $, and then outputs the result of $ \texttt{say\_something\_b()} $. After attacked, 
the invocation order of function $ \texttt{say\_something\_a()} $ and $ \texttt{say\_something\_b()} $ has been exchanged.

In the actual operation, the attacked program $app'$ executes normally without error. But the order of execution has changed. We can distinguish it by the different output rules of different ECall functions we have set before. At this point, the DDRC threat was completed.

\vspace{0.5em}
\par\noindent (b) OTM Threat Simulation

In this threat, we used the settings similar to those in the DDRC threat. We took the local-attestation project in sample code as the target. In this project, we can find the \texttt{secret\_data} in the untrusted area for exchanging information in the binary file by using the \textbf{\texttt{objdump}} and \textbf{\texttt{readelf}} tools. We added an xor operation to \texttt{secret\_data} with itself in the binary, which makes the \texttt{secret\_data} always 0. At this point, the Output Tampering threat is completed. For the Output Misrouting threat, after obtaining the location of \texttt{secret\_data}. We can redirect the variable to other functions by modifying the register corresponding to the value of the function arguments pushed into the stack. The Output Misrouting threat is completed.

\begin{figure*}
	\centering
	\includegraphics[width=7in]{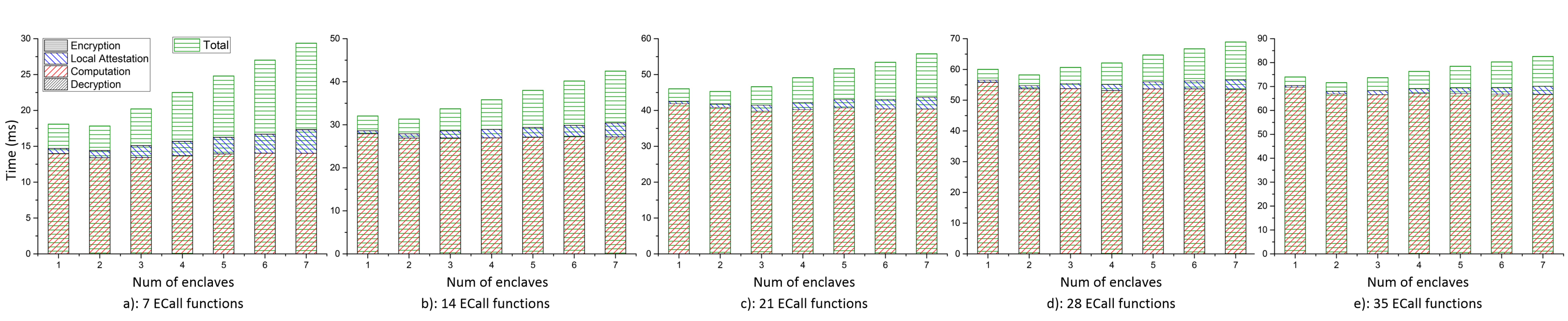}			
	\caption{Time cost of all the scheme-equipped experiments with different numbers of ECall functions}
	\label{figure14}
	\vspace{-1em}
\end{figure*}

\begin{figure*}
	\centering
	\includegraphics[width=7in]{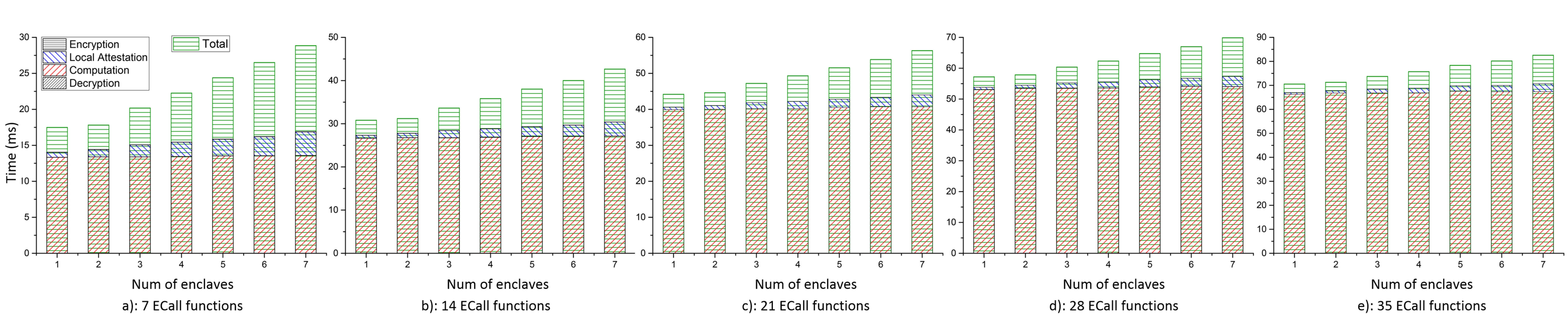}		
	\caption{Time cost of all the bare-metal experiments with different numbers of ECall functions}
	\label{figure15}
	\vspace{-1em}
\end{figure*}

\subsection{Correctness and Performance Evaluation} \label{sec6-B}

In this subsection, we present some experiments to further analyze the correctness, efficiency and capability of our scheme to verify the aforementioned theoretical analysis in Section \ref{sec5}. To ensure better accuracy, we performed each test 1000 times and recorded the average values of all the results. 
The max stack size and max heap size of each enclave program were set to \texttt{0x40000} and \texttt{0x100000}.

\subsubsection{Experiment Initialization}\label{subsubsec6-B}

Due to the limited size of Enclave Page Cache (EPC), assigning too many tasks to a single enclave leads to frequent memory page swapping and hence lower execution efficiency. For a specific computation task, the more enclaves are involved, the fewer tasks each single enclave gets. Hence, it can reduce the execution time and improve efficiency. But the increasing number of enclaves complicates the process of enclave creation and Local Attestation during the execution of a computation task, which leads to an increase to the whole execution time. Therefore, when deploying an SGX program, how to choose the proper number of enclaves to obtain the most efficient performance based on a specific computation task is a very challenging problem. In order to select a suitable number of enclaves for performance analysis of our verification scheme, we carried out the following experiments.

In order to make the experiment more clear and specific, we set each Enclave ECall implementation function to execute the following loop to simulate the real computation: \texttt{for(i=0;i<1000;i++)\{for(j=0;j<1000;j++)\{k=} \texttt{j+i;\}\}}


Considering that Local Attestation may wait for each other and cause unnecessary trouble to the later analysis, we carefully designed the attestation order between the enclaves, eliminating the mutual waiting between two enclaves.

We refer to the hybrid example in Fig. \ref{figure9} to set the number of ECall functions to 7, 14, 21, 28, 35, and set the total number of enclaves from 1 to 7. In each test, we uniformly deployed ECall functions in the enclaves. That is to say, each enclave has 3 ECall functions when there are 21 functions and 7 enclaves. We conducted comparative experiments to facilitate the analysis of the performance of our scheme. Therefore, a total of 35 sets of comparative experiments were conducted at the cloud. Measurements of the cloud experiments include decryption time, computation time, encryption time, Local Attestation time, and total time cost. The number of the ECall functions counted at the user is consistent with the cloud. The measurements of the experiments at the user include the time that the random number generation process spends, the time that the user spends to compute the hash code and performs verification, encryption, decryption, and total time cost.

\subsubsection{Correctness}In order to verify the correctness of our scheme, we compared the final hash codes obtained at the user and the cloud. Our experiments showed that the same hash code is obtained in all the test cases described in \ref{subsubsec6-B}.

\subsubsection{Performance Evaluation}

Our experimental results are shown in Figs. \ref{figure14} and \ref{figure15}. For each experiment in the figures, encryption, Local Attestation, computation, and decryption cost are shown in a stack column graph, i.e., each part is above the adjacent part. For the total time cost, we can compare it with the graph of the former four parts. We set the starting position of the total time cost to 0, that is, it is not a stack column graph with the former four parts. In order to facilitate the analysis of the time cost in each comparative experiment, we set the ordinates of different scale values for the number of different ECall functions. We present the annotations for each in Figs. \ref{figure14} and \ref{figure15}, respectively.

From Figs. \ref{figure14} and \ref{figure15}, we can see that with the same number of ECall functions, as the number of enclaves increases, the time required is roughly incremental. But for Fig. \ref{figure14}, we find that regardless of the number of ECall functions, when the number of enclaves is 2, both the computation and the total time are the least. Next, although the number of enclaves is not the least when the number of enclaves is 2 in Fig. \ref{figure15}, in the average sense, the time cost of 2 enclaves is only 1.26\% more than that of a single enclave. Therefore, for our execution plan, the enclave program performs best when it is 2 enclaves. Therefore, we chose the comparative experiments with 2 enclaves to evaluate the performance of our scheme.

\begin{figure}
	\centering
	\includegraphics[width=3.5in]{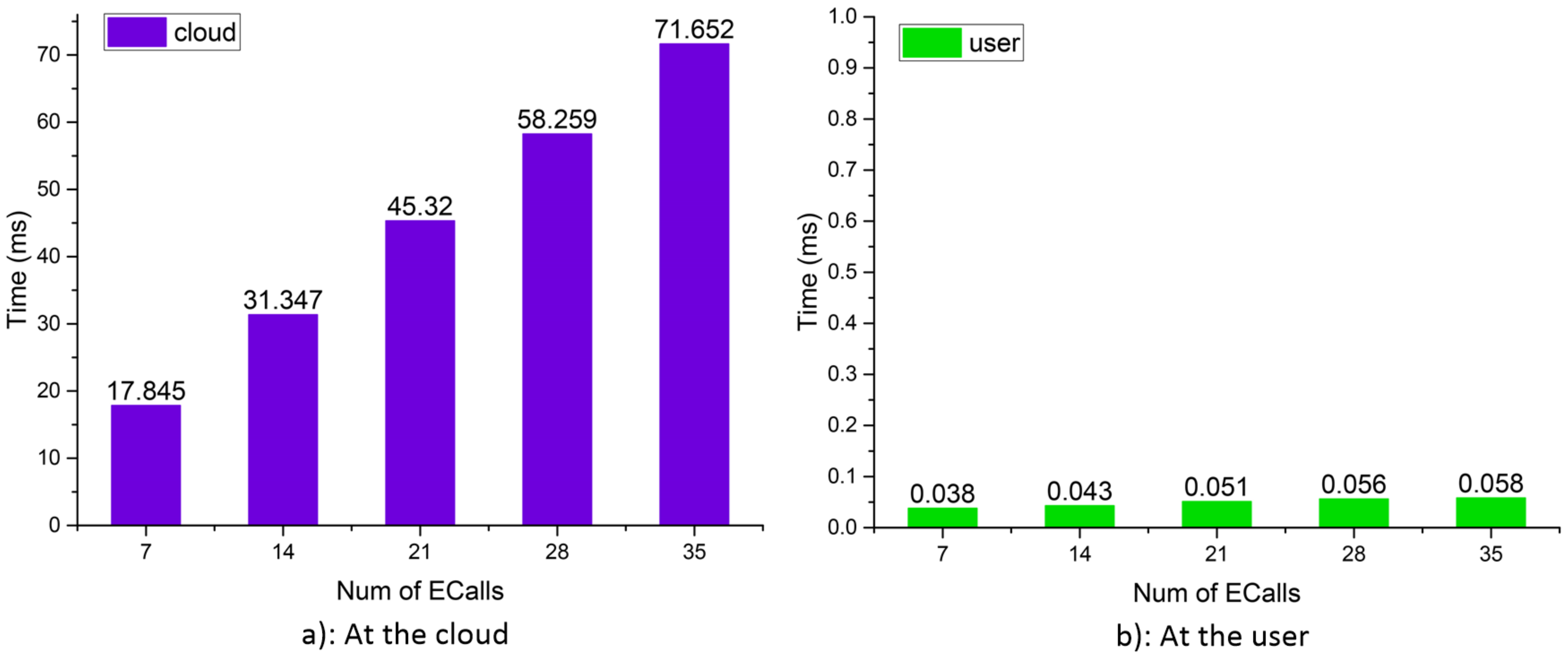}	
	\caption{Time cost comparison between cloud and user: a) Total time cost of all the scheme-equipped experiments with 2 enclaves at the cloud; b) Total time cost of all the experiments at the user}
	\label{figure16}
	\vspace{-1em}
\end{figure}

The computational costs of the cloud and the user are shown in Fig. \ref{figure16}. The left part shows the total time cost of the experiments with 2 enclaves as the number of ECall functions increases. The right part shows the total time cost at the user. As we can see, the computational cost at the user is less than 0.1 ms, which is suitable for the user device with limited resources.

\begin{figure}
	\centering
	\includegraphics[width=3.5in]{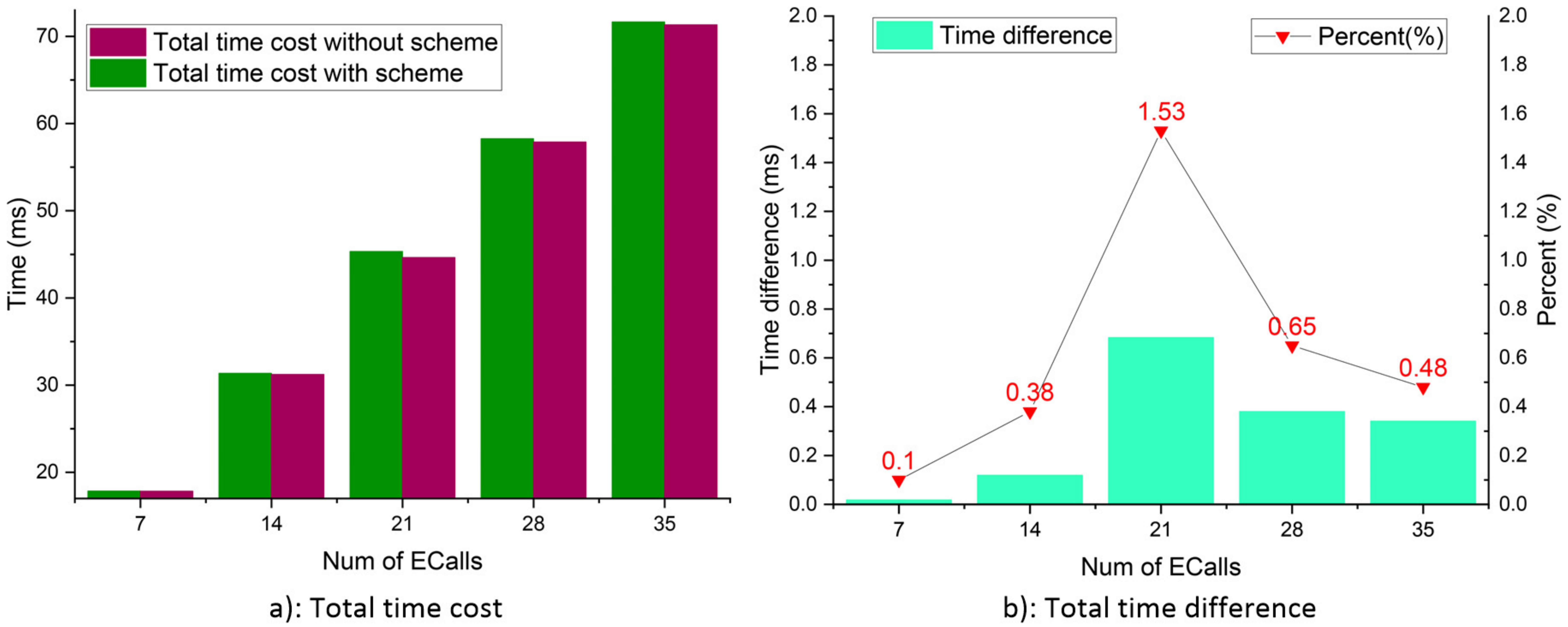}					
	\caption{Time cost difference between scheme-equipped experiment and bare-metal experiment at the cloud: a) Total time cost with 2 enclaves; b) Total time difference with 2 enclaves}
	\label{figure13}
\end{figure}

The total time cost difference between scheme-equipped experiment and bare-metal experiment with 2 enclaves is shown in Fig. \ref{figure13}. As the number of ECalls increases, there is a comparison of the total time cost between bare-metal and scheme-equipped experiments in the left part of the figure. In order to show the time difference more clearly, we show the corresponding time difference in the right part. We can see that all the time differences are less than 1ms, which indicates that our scheme has a very low computational overload to the SGX bare-metal computation. At the same time, we also show the percentage of the time difference in the form of a line chart on the right part. The computation formula of the percentage of the time difference is $( time\_with\_scheme - time\_without\_scheme) / time\_without\_time * 100\%$. The right axis of the right graph is its corresponding coordinate axis, and the range is 2\%. So we can easily see from Fig. \ref{figure13} that the overhead of the solution is extremely low, which indicates that our scheme can be well accepted by the cloud service providers.

\subsubsection{Comparison with Cryptography-Based Scheme}

We further compared our SGX-based verifiable computation scheme with a cryptography-based verifiable computation one \cite{yu2019verifiable}. In this experiment, we set that FHE public key and secret key are 440 bytes and 464 bytes, respectively.

\begin{table}
\caption{Comparison with cryptography-based scheme \cite{yu2019verifiable}}
\begin{center}
\begin{tabular}{|c|c|l|r|}
	\hline
	\multicolumn{1}{|l|}{}                          & \textbf{Phase}                               & \textbf{Entity} & \textbf{Time(ms)} \\ \hline
	\multirow{6}{*}{Scheme in \cite{yu2019verifiable}}               & \multirow{4}{*}{Computation phase}  & DP$^{\mathrm{\star}}$     & 183.118  \\ \cline{3-4} 
	&                                     & CSP$^{\mathrm{\ast}}$    & 243.541  \\ \cline{3-4} 
	&                                     & RP$^{\mathrm{\circ}}$     & 10.179   \\ \cline{3-4} 
	&                                     & TA$^{\mathrm{\dagger}}$     & 298.675  \\ \cline{2-4} 
	& \multirow{2}{*}{Verification phase} & DP$^{\mathrm{\star}}$     & 364.394  \\ \cline{3-4} 
	&                                     & PAP$^{\mathrm{\ddagger}}$    & 10.980   \\ \hline
	\multicolumn{1}{|l|}{\multirow{3}{*}{Our work}} & \multirow{2}{*}{Computation phase}  & Cloud  & 4.142    \\ \cline{3-4} 
	\multicolumn{1}{|l|}{}                          &                                     & User   & 0.007    \\ \cline{2-4} 
	\multicolumn{1}{|l|}{}                          & Verification phase                  & User   & 0.021    \\ \hline
	\multicolumn{4}{l}{$^{\mathrm{\star}}$DP: Data Provider. $^{\mathrm{\ast}}$CSP: Cloud Service Provider.}\\ 
	\multicolumn{4}{l}{$^{\mathrm{\circ}}$RP: Requesting Party. $^{\mathrm{\dagger}}$TA: Trusted Authenticator.}\\ 
	\multicolumn{4}{l}{$^{\mathrm{\ddagger}}$PAP: Public Auditor Proxy.}\\ 
\end{tabular}
\label{tab7}
\end{center}
\vspace{-1em}
\end{table}


The time costs of all the entities involved in the two schemes are listed in Table \ref{tab7}. We can see that in both the computation and verification phase, the scheme \cite{yu2019verifiable} is more time-consuming than ours. Especially, the low computation overhead of our scheme at the user side shows much superiority over that of the scheme \cite{yu2019verifiable} at DP and RP, which makes it extremely suitable for users with limited resources. Our scheme also greatly reduces the computation overhead at the cloud and gets rid of the dependence on any trusted third-party servers. In general, our proposed scheme has a simple system structure and high computation efficiency, which is very suitable for being applied to various real-world scenarios.

\section{Conclusion} \label{sec7}

In this paper, we proposed an efficient and secure verifiable computation scheme for Intel SGX, which can resist two types of threats (DDRC threat and OTM threat) and overcome the shortcomings of cryptography-based verifiable computation schemes. We demonstrated the validity of the two new threats towards the program constructed by Rust SGX SDK. Moreover, we analyzed the security and computational complexity of our proposed scheme, and carried out comprehensive experiments to show its correctness and high efficiency for resisting the aforementioned threats.

%

%
%
%
%

\bibliographystyle{IEEEtran}
\bibliography{IEEEabrv,reference}


\end{document}